\documentclass{aa}
\usepackage{graphicx,times}
\usepackage{natbib}
\bibpunct{(}{)}{;}{a}{}{,}
\newcommand{\aap}{A\&A\ }

\def\ch{{\it Chandra}}

\def\Halpha{\ifmmode {\rm H}\alpha \else H$\alpha$\fi}
\def\Hbeta{\ifmmode {\rm H}\beta \else H$\beta$\fi}
\def\Hgamma{\ifmmode {\rm H}\gamma \else H$\gamma$\fi}
\def\Hdelta{\ifmmode {\rm H}\delta \else H$\delta$\fi}
\def\Lya{\ifmmode {\rm Ly}\alpha \else Ly$\alpha$\fi}
\def\Lyb{\ifmmode {\rm Ly}\beta \else Ly$\beta$\fi}
\def\Lyg{\ifmmode {\rm Ly}\beta \else Ly$\gamma$\fi}

\def\hi{H\,{\sc i}}

\def\ciii{\ifmmode {\rm C}\,{\sc iii} \else C\,{\sc iii}\fi}
\def\civ{\ifmmode {\rm C}\,{\sc iv} \else C\,{\sc iv}\fi}
\def\cv{\ifmmode {\rm C}\,{\sc v} \else C\,{\sc v}\fi}
\def\cvi{\ifmmode {\rm C}\,{\sc vi} \else C\,{\sc vi}\fi}

\def\o5007{[O\,{\sc iii}]\,$\lambda5007$}

\def\ov{O\,{\sc v}}

\def\ovii{O\,{\sc vii}}
\def\oviii{O\,{\sc viii}}

\def\o{\o}
\def\kms{{km\,s$^{-1}$}}

\begin{document}

\title{ Comments on the manuscript: ``The weak absorbing outflow in
AGN Mrk 279: evidence of super-solar metal abundances''
astro-ph/0611578 by Fields et al.}

\author{Nahum Arav\inst{1}
\and Jelle Kaastra\inst{2}
\and Elisa Costantini\inst{2,3} }

%\offprints{}
%\mail{}

\institute{CASA, University of Colorado, 389 UCB, Boulder, CO 80309-0389, USA
              \and
		SRON Netherlands Institute for Space Research
              Sorbonnelaan 2, 3584 CA Utrecht, The Nether\-lands
	   \and
              Astronomical Institute, Utrecht University, 
                P.O. Box 80000, 3508 TA, Utrecht, The Netherlands
}

\date{}

%\authorrunning{ }
%\titlerunning{ }

\abstract{ A recent manuscript posted on astro-ph (0611578) by Fields
et al. (hereafter F06) reports evidence of supersolar metal abundances
in Mrk 279 by analyzing its Chandra LETGS X-ray spectrum.  We point
out that it is impossible in principle to obtain direct metal
abundances from these X-ray data, since there is no handle on the
amount of hydrogen column density.  If F06 would have lowered their C,
N, O and Fe abundance by a factor of ten and increased the hydrogen
column density by a factor of ten, they would have obtained an almost
identical fit with subsolar metalicity.  F06 find support for their
supersolar metal abundances from a cursory analysis of the UV data
from the same Mrk 279 campaign.  We point out that F06 included in
that analysis portions of the UV trough that are known to arise from
gas unrelated to the outflow, which weakens the support from the UV
data.  A detailed analysis of the Chandra LETGS X-ray spectrum was
accepted for publication in A\&A on Sept 14 2006 (Costantini et al
2006; hereafter C06) and posted on astro-ph on the same date.  F06
ignore most of this published analysis while duplicating the finding
of two ionization components with similar parameters to the ones found
by C06. Finally, we note that it is possible to derive accurate
abundances from the UV data set of this object.  We already published
these findings in a conference precedings and have submitted the
relevant manuscript to ApJ.  We find that relative to solar the
abundances in the Mrk~279 outflow are (linear scaling): 
carbon 2.2$\pm0.7$ , nitrogen 3.5$\pm1.1$ and oxygen 1.6$\pm0.8$.   
}
\maketitle

\section{Introduction}

We comment on the Fields et al. paper entitled ``The weak absorbing
outflow in AGN Mrk~279: evidence of supersolar metal abundances" that
recently appeared on astro-ph (astro-ph/0611578; hereafter F06).  This
paper analyses the data from a deep simultaneous \ch
-LETGS/HST-STIS/FUSE spectroscopic campaign on the AGN outflow seen in
this object.  We, the PI team, published four refereed papers on this
data set (Kaastra et al. 2004; Gabel et al. 2005; Arav et al. 2005;
Costantini et al. 2006) and submitted another manuscript to ApJ a few
months ago (Arav et al. 2006b). We find several problems with the F06
analysis and their reference to our previous work on this data set. In
this comment we elaborate on these points.

\section{Fitting the X-ray data}\label{par:xray}

{\em a)} F06 reports that a single power-law plus Galactic absorption provides a
good fit to the LETGS data between 10-50 \AA\ (see their
Sect. 3.1). In contrast, in C06 we find that two components are required:
in addition to a power-law ($\Gamma\sim 2$) the spectrum requires a
soft excess with a break energy above 10\,\AA\ (See Tab.~1 of
C06). {\em b)} Moreover, around energies of crucial oxygen ion
transitions (\ovii\ triplet, \oviii\ \Lya) we found evidence of broad
excess emission (first shown by Kaastra et al. (2004) and accurately
modeled in C06).  Not taking those gaussian-shaped emission components
into account can lead the spectral-fit astray and compromises the
calculation of the column density for \ovii\ and \ov. {\em c)}
Finally, F06 assume throughout the paper an arbitrary velocity
dispersion of 100\,\kms, which may cause additional uncertainties in
fitting the data and for column densities estimates. Using the
information on both the most prominent X-ray ions fitting and the UV
absorption lines, C06 adopted instead 50\,\kms.

The treatment of the errors is quite vague throughout the F06 paper. In
particular,  there are no quoted errors on the non-solar abundances, which is
the major topic of the paper.  It seems remarkable that a reduced $\chi^2$
less than unity applies for all the models tested.\\

Another minor point on the X-ray analysis is related to the Kaastra et al. 2004
paper, where we report the tentative detection of the \ov* absorption line,
produced from a meta-stable level. F06  confuses the \ov * line with the \ov\
resonant line (Sect.~1 of F06), whose detection is solid.

\section{Inability to obtain direct metal abundances from these X-ray data}

F06 report evidence for supersolar metal abundances in Mrk~279 by
analyzing its \ch-LETGS X-ray spectrum. They conclude that a
photoionization model with abundances of 2$_\odot$ for carbon, 5$_\odot$ for
nitrogen, 7$_\odot$ for iron, and 8$_\odot$  for oxygen, gives a better fit to the
\ch-LETGS data and therefore infer super-solar abundances for this
object.  However, we point out  that it is impossible in
principle to obtain direct metal abundances from these X-ray data,
since there is no handle on the amount of hydrogen column density.

\begin{figure}
\resizebox{\hsize}{!}{\includegraphics[angle=90]{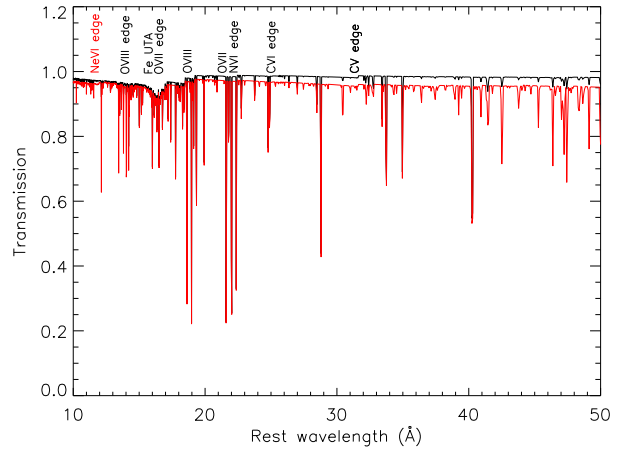}}
\caption{Black line: F06 best fit transmission spectrum as a function
of the rest wavelength (our model 1, see text). This model has
overabundances for C, N, O and Fe. Red line: model 2, with ten times
larger total column density but ten times lower abundances of C, N, O
and Fe. The models are very similar despite the large abundance
changes, where the EW of the C, N, O, and Fe lines are identical for 
both simulations.
The 1--3\% discrepancies between the two
curves are entirely due to the enhanced column density of the trace elements
such as Ne, Mg, S, and Si (see e.g. the neon inner shell edge labeled in model
2) whose abundances were not changed}
\label{f:fig1}
\end{figure}

In Fig.~\ref{f:fig1} the simulated best-fit transmission spectrum for
the super-solar metalicity model of F06 (their model 3) is displayed
by the black curve (our model 1). The simulation was performed using
SPEX\footnote{http://www.sron.nl/divisions/hea/spex/version2.0/release/index.html}
(ver.~2.0).  The simulation includes two ionized components with
Log$U$=--0.73, Log$N_{\rm H}$=19.66\,cm$^{-2}$ and Log$U$=0.93,
Log$N_{\rm H}$=19.50\,cm$^{-2}$, respectively, and a line velocity
dispersion of 100\,\kms. The abundances of C, N, O, and Fe were set to
2$_\odot$, 5$_\odot$, 8$_\odot$, and 7$_\odot$, respectively, as in
the F06 best fit. Note the prominent bending caused by the
enhanced Fe and O abundances at around 15-17\,\AA, as to mimic a break
in the continuum spectrum. Therefore, at least a part of the required
overabundances in F06 may be indeed the result of the incomplete
continuum fitting discussed in  point a) of \S~\ref{par:xray}.

The red curve in Fig.~\ref{f:fig1} shows the same ionization
components, but with C, N, O and Fe abundances set to 1/10 of the
values for model 1, while the total $N_{\rm H}$ (and coupled to it,
other trace elements) is enhanced by a factor of 10 (our
model~2). Therefore, the total ionic column densities of C, N, O and
Fe are the same in model 1 and 2.  The small changes between the two
models are caused by other elements (e.g. Ne, Mg, Si, S etc.), which
are responsible for the enhanced opacity in the 10-12\AA\ region and
above 20 \AA. As hydrogen is almost fully ionized, it does not
contribute significantly to the X-ray opacity.

\begin{figure}
\resizebox{\hsize}{!}{\includegraphics[angle=90]{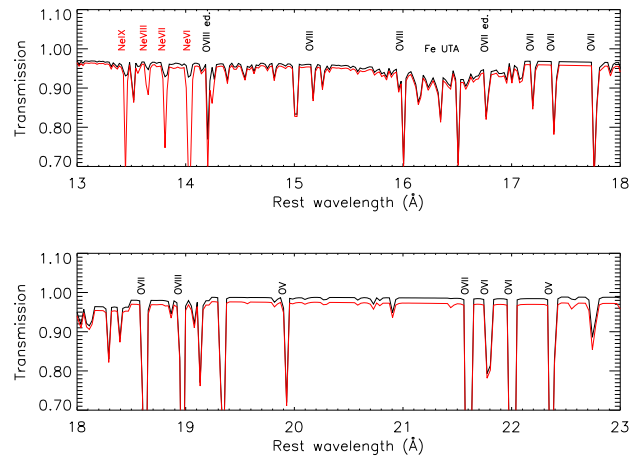}}
\caption{Detail of Fig.~\ref{f:fig1} in the 14-18 \AA\ and 18-23 \AA\
regions.  The color convention follows the same definition as in
Fig.~\ref{f:fig1}.  It is evident that the iron and oxygen line profiles
are identical for the two models, in spite of the ten times difference
in their abundances.  The Neon lines around 13--14\AA\ show dramatic
differences since we did not change the Neon abundance while
increasing the total column density by a factor of ten.}
\label{f:fig2}
\end{figure}

In Fig.~\ref{f:fig2} a detail of the 14-18\AA\ and 18-23\AA\ region is
shown.  The line profiles of oxygen and iron are the same
for both models, despite the fact that the abundances of those elements are a
factor ten lower.

In summary, virtually identical column densities for the observed C,
N, O, and Fe troughs are obtained with a model that has exactly 1/10
of the F06 abundances (i.e., a subsolar metalicity model) and 10 times
the total hydrogen column density.  This is especially true since the
low total column density in this object does not allow for strong
bound-free edges to develop, which might affect the global shape of
the spectrum. Therefore, we find the claim of super-solar abundances
based on the \ch-LETGS data untenable.

\section{Problems with the supportive UV analysis}

F06 find support for their supersolar metal abundances from a cursory analysis
of the UV data from the same Mrk~279 campaign. However, F06 used the reported
ionic column density of Gabel et al. (2005) for the combined 2+2a and 4a
components.  As pointed out in Scott et al. (2004, Sect.~3.4) and Gabel et al.
(2005) much of the absorption feature associated with component 4 (including 4a)
is due to gas that is unrelated to the intrinsic outflow in Mrk~279. Inclusion
of measurements from that gas cast doubt on the validity of the F06
photoionization plots (their figs. 4 and 5) and affects their conclusions from
the UV data.  Furthermore, F06 did not attempt to solve for the ionization
equilibrium and abundances of the UV absorber. They merely claimed that their
X-ray photoionization solution for the supersolar abundances case is a better
fit for the UV data than their solar abundances X-ray photoionization solution. 
But as pointed out above their X-ray fit could be achieved with a subsolar
metalicity model.

\section{Ignoring and referencing published results}

The paper by F06 was only recently submitted to ApJ, while C06 was
accepted by A\&A more than two months earlier (Sept 14 2006).  It is
unfortunate that in their Introduction F06 ignores C06 and only
mention an out-of-date, un-refereed conference proceedings that
appeared in 2005 (Costantini et al. 2005). Due to the above time-line
we are also puzzled by the F06 sentence (\S~4.3) regarding the C06
work: ``Their new results agree with ours...'' Proper referencing
should have been ``Our results agree with C06...''  In addition F06
seems to ignore much of the results on the warm absorber that were
presented in C06: For example The shape of the underlying continuum
and the detected X-ray broad emission lines have direct relevance to
the analysis of this warm absorber.

\section{Can abundances be determined from this data set?} 

It is possible to derive accurate abundances from the UV data set of
this object since these data include the crucial \hi\ absorption
troughs.  We presented these findings several conferences, published
them in a conference proceedings (Arav 2006a) and submitted a
manuscript to ApJ (Arav et al. 2006b) with these results, which passed
through one refereeing cycle. In that paper, we determine the chemical
abundances by using a careful velocity-dependent analysis of portions
of the trough that are unambiguously associated with the outflow.  We
find that relative to solar the abundances in the Mrk~279 outflow are
(linear scaling): carbon 2.2$\pm$0.7, nitrogen 3.5$\pm$1.1 and oxygen
1.6$\pm$0.8. We will post this detailed analysis as soon as the paper
is accepted.

\end{document}